# Precision Through Play: Enhancing Fundamental Laparoscopic Skill Acquisition Via Gamified XR and Pseudo-Haptics

Bita koohestani[1,2], Ashkan Rezaei[1,2], Azam Bastanfard[1,2]*

[1] Department of Computer Engineering, Ka.C., Islamic Azad University, Karaj, Iran.

[2] The Research Center for Interactive Technologies and Extended Reality, Islamic Azad University Karaj Branch

*Corresponding author(s). E-mail(s): Bastanfard@iau.ac.ir
Contributing authors: bita.koohestani@iau.ir; ashkan.rezaei0702@iau.ir;

**Abstract**

Extended Reality (XR) technologies have shown potential to improve laparoscopic skills acquisition. Among fundamental skills in laparoscopic surgery is accurate instrument handling. Appropriate instrument alignment and a perpendicular path ensure movement efficiency and improved performance. This work introduces an interactive XR platform designed to support an immersive, engaging, and accessible environment that facilitates continuous practice. This platform incorporates a gamification layer enriched by pseudo-haptics that generate a motivational feedback loop, allowing trainees to continuously practice instrument manipulation and angle control. The platform monitors instrument orientation, computes angular deviation from an ideal reference vector, and assigns scores based on user performance. A point-based scoring component reinforces correct technique and penalizes deviations in tool alignment beyond predefined thresholds. Furthermore, a pseudo-haptic feedback mechanism is integrated to provide real-time tactile feedback when deviation exceeds tolerance. Combined with low-cost hardware and accessible deployment, the framework aims to increase training scalability, engagement, and improve skill acquisition. Preliminary evaluations indicate that integrating gamified feedback significantly improves precision, reduces angular variance, and encourages sustained practice.

**Keywords:** Extended Reality, Pseudo-Haptics, Gamification, Laparoscopic Training, Surgical Education

## 1 Introduction

Minimally invasive surgery (MIS) gives rise to fundamental alterations in the surgeon's sensory environment (Okamura, 2009). Loss of direct tactile perception (van der Mei-Jden and Schijven, 2009) and reliance on two-dimensional line-of-sight visualizations deprive the surgeon of natural cues that guide instrument navigation. By forcing the surgeon to operate through small incisions, MIS converts typical three-dimensional (3D) perception into constrained two-dimensional (2D) visual feedback, increasing cognitive load and reducing intuitive guidance (Chen et al., 2021). Due to this dissociation, trainees attempting to master surgical skills such as hand-eye coordination, instrument alignment, and depth perception face several complications (Vine et al., 2012). As a result, the early stages of laparoscopic training remain challenging and often inefficient without structured, timely guidance (Takbiri et al., 2019).

Among the most essential competencies required in laparoscopic surgery, proper instrument angling and alignment are of critical importance (Kunert et al., 2020). An appropriate insertion angle of the trocar with a perpendicular path minimizes tissue trauma, optimizes laparoscopic camera visibility, and improves intraoperative visualization (Paasch et al., 2024). Despite the critical nature of this skill, trocar angle control is frequently under-emphasized in early-stage training curricula, where instruction is often informal and practice unstandardized. Traditional training approaches such as static box trainers or benchtop simulators offer a basic environment for initial familiarization, but typically lack mechanisms for precise angle monitoring or automated corrective feedback (Abinaya and Manivannan, 2024). Their effectiveness further relies heavily on direct mentoring from experienced surgeons and access to specialized simulation spaces and hardware, both of which are limited in many regions (F et al., 2015). This results in a critical training gap where fundamental spatial and alignment skills are not adequately reinforced through immediate, objective feedback. Recent advances in Extended Reality (XR) technologies present promising alternatives to overcome these persistent limitations. By merging the real-world physical environment with virtual overlays, XR can provide immersive, simulated training experiences on widely available devices (Seymour et al., 2002). Unlike conventional VR systems, XR platforms can leverage smartphones, tablets, or lightweight head-mounted displays (HMDs), enabling broader access to high-quality surgical training without high costs or complex infrastructure. These platforms introduce instrument motion visualizations, display optimal trajectories, and provide feedback mechanisms that monitor user performance in real time (Bui et al., 2025). However, while XR addresses limitations of accessibility and visualization, most implementations still lack effective sensory feedback mechanisms and motivational reinforcement, which are critical for sustained learning.

Pseudo-haptic feedback has been introduced as a practical, hardware-independent means of simulating tactile sensations in XR environments. By manipulating controlled visual cues such as motion distortion, delayed response, and altered tool dynamics, and combining them with subtle vibration patterns, pseudo-haptics creates an illusion of resistance or force feedback without specialized hardware (L´ecuyer, 2009). This trick on the user's perception exploits the brain's merging of vision and proprioception, enabling trainees to perceive corrections or mistakes through simulated sensory feedback. In context of laparoscopic skill training, such mechanisms can signal improper trocar alignment or angle deviation by providing brief vibration pulses (Hirao et al., 2022) or altering tool responses (Li et al., 2014). These strategies offer a cost-effective means of reinforcing performance correction while remaining compatible with mobile XR devices (Hirao et al., 2023).

At the same time, educational research has brought the motivational value of gamification to light, highlighting its importance in medical skill acquisition (Turchet et al., 2025). Elements such as scoring systems, performance progression, and real-time feedback loops encourage repeated practice and enhance learner engagement (Samadzadeh Tabrizi et al., 2025). When fully integrated, gamification transforms repetitive procedural practice into structured, goal-oriented challenges that engage learners' attention and support long-term retention (Verdaasdonk et al., 2009). Despite its potential, many existing XR training platforms overlook these motivational design principles, limiting their potential for sustained, continuous training. Consequently, there remains a gap for platforms that integrate engaging game-design ele-ments and motivational feedback to drive educational reinforcement (Krishnamurthy et al., 2022).

The proposed method addresses these gaps by presenting an XR platform that combines a gamified scoring module and pseudo-haptic guidance. The platform continuously monitors instrument angle and trajectory, providing multi-sensory cues when deviations from the acceptable trocar path occur. Deviations from acceptable trocar angles trigger point deductions and subtle vibration pulses, while proper handling is rewarded, creating a structured learning loop and reinforcing proper technique through immediate reward. By combining pseudo-haptic interaction and gamified learning on affordable consumer devices, this platform aims to close existing training gaps by providing accessible, high-fidelity, and pedagogically engaging laparoscopic skill development—particularly for learners in resource-constrained environments.

In the following sections, this paper first reviews relevant literature on laparoscopic training, Extended Reality applications, and the use of pseudo-haptic feedback and gamification in surgical education. This background sets the stage for the proposed solution—an XR-based platform that combines real-time angle tracking with multi-sensory feedback and motivational gamified elements. The subsequent section outlines the experimental framework used to evaluate training effectiveness, user engagement, and performance outcomes. The results and discussion highlight how the platform enhances both engagement and skill acquisition in minimally invasive surgery training. Finally, the paper concludes with reflections on the implications of this work and suggests further expanding the reach and pedagogical impact of XR-based surgical training solutions.

## 2 Background and Related Work

### 2.1 Scope and Challenges

The shift to MIS from open surgery has reformed surgical education by introducing distinct constraints such as ergonomic and perception challenges on the surgeon. Despite significant benefits offered by MIS-reduced trauma, shorter recovery time, and lower infection rates, it imposes the ”fulcrum effect”, where instrument tips move in the opposite direction of the handle (Tokunaga et al., 2012). This effect, coupled with reduced field of view and a loss of direct tactile feedback (Colan et al., 2022), introduces a steep learning curve and increases cognitive load. As a result, the conventional apprenticeship model (”see one, do one, teach one”) (Halsted, 1904) is no longer sufficient for early laparoscopic training.

Current training standards focus on simulation-based education (SBE) to allow trainees to acquire skill proficiency in a risk-free environment. Traditional box trainers deliver valuable haptic fidelity and instrument handling experience but rarely provide objective assessment metrics (Van Sickle et al., 2011). High-fidelity Virtual Reality (VR) simulators offer detailed metrics and capture detailed performance data but remain resource-intensive and are frequently confined to dedicated simulation cen-ters (Munz et al., 2011; Scott et al., 2019). There is, therefore, a pressing need for training solutions that balance tactile realism, real-time metric tracking, and wide accessibility (Mulla et al., 2019).

### 2.2 XR as a Surgical Education Tool

Extended Reality encompasses Augmented Reality (AR), Virtual Reality (VR), and Mixed Reality (MR), offering an adaptable platform for bridging the gap between immersive visualizations, precision, and accessibility (Barsom et al., 2016). These technologies deliver controlled, risk-free environments grounded in the principles of the Virtual Physiological Human (Magnenat-Thalmann and Thalmann, 2004), allowing trainees to rehearse without patient risk and limitations of cadaver-based or apprenticeship learning (Kunkler, 2006). Existing systems have demonstrated substantial promise to improve hand-eye coordination, depth perception, and procedural accuracy (Bric et al., 2009). Such skills are fundamental in laparoscopic surgery, as limited field of vision and tactile cues impose a steep learning curve.

VR simulators such as LapSim and LapMentor were among the earliest systems to introduce structured performance metrics, enabling evaluation of path lengths, angular precision, and tissue handling errors (Munz et al., 2011). More advanced solutions, including LapVR, incorporate true force-feedback and biomechanically accurate organ models to provide enhanced fidelity (Seymour et al., 2002; Maurel et al., 1998). Although these simulators have shown strong evidence of skill transfer to clinical practice, their reliance on specialized workstations, motion-tracking hardware, and high acquisition costs restricts deployment to well-funded academic centers.

AR-based approaches have been introduced to address these accessibility constraints. By overlaying virtual guidance cues on real physical instruments, AR supports laparoscopic skill acquisition (Barsom et al., 2016). Existing AR-based platforms, such as Touch Surgery, have shown substantial promise by integrating procedural walk-throughs and cognitive practice on lightweight devices (Seymour and Gallagher, 2012). Nevertheless, these platforms typically lack real instrument tracking, objective motor-skill evaluation, or tactile cues. Most AR solutions remain single-use, visually guided systems that do not replicate the spatial or sensory load of real laparoscopic tasks.

XR lies at the intersection of these methods, offering a unified continuum that blends immersive visualization with physical interaction grounded in the real world.

For laparoscopic training, XR enables the integration of real instrument handling, immersive context-aware overlays, and dynamic feedback while maintaining hardware portability. This combination provides a more flexible and scalable alternative to traditional simulators, supporting both physical instrument manipulation and virtual instructional enhancements. As a result, XR-based platforms are increasingly viewed as promising candidates for democratizing early-stage laparoscopic education—particularly when optimized for consumer-grade hardware and enhanced with features like gamification and pseudo-haptics (Gurusamy and Aggarwal, 2020).

## 2.3 Gamification in Medical Education

Gamification integrates game-design elements-such as scoring, progression, and immediate feedback- into training environments to increase motivation and encourage engagement through continuous practice (Takbiri et al., 2023; Hamari et al., 2014). In the context of skill learning, gamification promotes engagement, encourages repetition, and can improve retention when the game mechanics' aim is parallel to learning objectives(Graafland et al., 2012). For laparoscopic training, effective gamification must align rewards and penalties with relevant performance metrics such as instrument angle stability and path smoothness to avoid encouraging superficial or counterproductive actions.

Research indicates that gamified elements such as scoring systems, leaderboards, and immediate error feedback can transform repetitive drilling into ”deliberate practice” (Samadzadeh Tabrizi et al., 2025). In laparoscopic training, studies have observed that competitive scoring encourages trainees to perform more voluntary repetitions than non-gamified control groups (Verdaasdonk et al., 2009). Despite these benefits, few XR surgical systems combine surgical education and structured gamified scoring based on performance metrics like instrument angle control, ensuring skill retention.

## 2.4 Pseudo-Haptics and Sensory Substitution

One of the most persistent limitations of portable and low-cost simulation is the absence of authentic force feedback. Traditional haptic devices involve complex and bulky hardware, rendering them expensive and nonportable. To overcome this limitation, researchers have turned to pseudo-haptics and sensory substitution (L´ecuyer, 2004).

Pseudo-haptics typically refers to the generation of a haptic sensation such as friction or stiffness through visual manipulation, tricking the brain into perceiving physical resistance. In sensory substitution, one sensory modality is replaced or augmented by another (Swindells and Mon-Williams, 2000). In practice, pseudo-haptic mappings may include motion scaling, dynamic damping of the virtual instrument, depth distortion, or vibration patterns tied to error magnitude (Min and Kim, 2012). In the context of this work, by mapping angular deviation to distinct vibration patterns, a system can alert the user to errors without requiring heavy robotic arms. This approach allows for lightweight, wearable or handheld feedback mechanisms that maintain the portability of the training system while restoring a crucial loop of non-visual information. Importantly, the effectiveness of pseudo-haptic methods depends on latency, display fidelity, and individual perceptions; thus, system design must prioritize responsive rendering and user-specific tuning.

### 2.5 Convergence and Existing Gaps

Prior work has extensively explored XR, gamification, and haptic technologies, yet fully integrated, low-cost platforms that combine all three remain rare. High-end commercial simulators successfully integrate immersive visualization, gamified exercises, and active haptics, but their cost and requirement for specialized centers hinder broad adoption. Mobile apps and simple AR prototypes provide accessibility and gamified interaction but often rely solely on visual cues, increasing visual cognitive load and limiting multisensory learning. Therefore, there is a distinct lack of ”middle-ground” solutions that utilize consumer-grade XR hardware to track physical instruments while providing multi-sensory feedback. Most existing low-cost solutions rely solely on visual cues, which increases visual cognitive load.

This work addresses that gap by proposing a consumer-grade XR platform that synthesizes instrument-level tracking, graded pseudo-haptic feedback, and an objective gamified scoring model to promote deliberate, measurable practice in trocar angle control—thereby delivering a practical, scalable training loop that current approaches do not offer.

## 3 METHODOLOGY

The proposed method is an XR simulation platform designed to facilitate laparoscopic surgical training through immersive visualization, real-time performance monitoring, and multimodal feedback. The methodology reflects the pedagogical challenges identified earlier—lack of continuous feedback, limited accessibility of high-fidelity training, and the absence of structured, gamified skill reinforcement. This section details the architecture, interaction mechanics, angular deviation estimation, and the gamified assessment and scoring. Figure 3 outlines the architecture of this platform.

### 3.1 System Architecture and XR Framework

The simulator utilizes the Unity Engine as the core platform, which enables real-time 3D rendering, scene management, and multimodal user interaction control (Unity Technologies, 2025b). The architecture leverages the Unity Engine and AR Foundation to unify ARCore and ARKit functionalities (Unity Technologies, 2025a; Google Developers, 2025; Apple Inc., 2025). The XR Core relies on AR Foundation, providing a unified abstraction layer over platform-specific SDKs through the XR Plugin Management Stack. The framework supports essential functionalities, including 6-DOF camera tracking, environmental mapping, and stable world-anchored registration of virtual surgical elements. This approach addresses accessibility challenges discussed earlier by situating the training environment within the trainee’s real physical space.

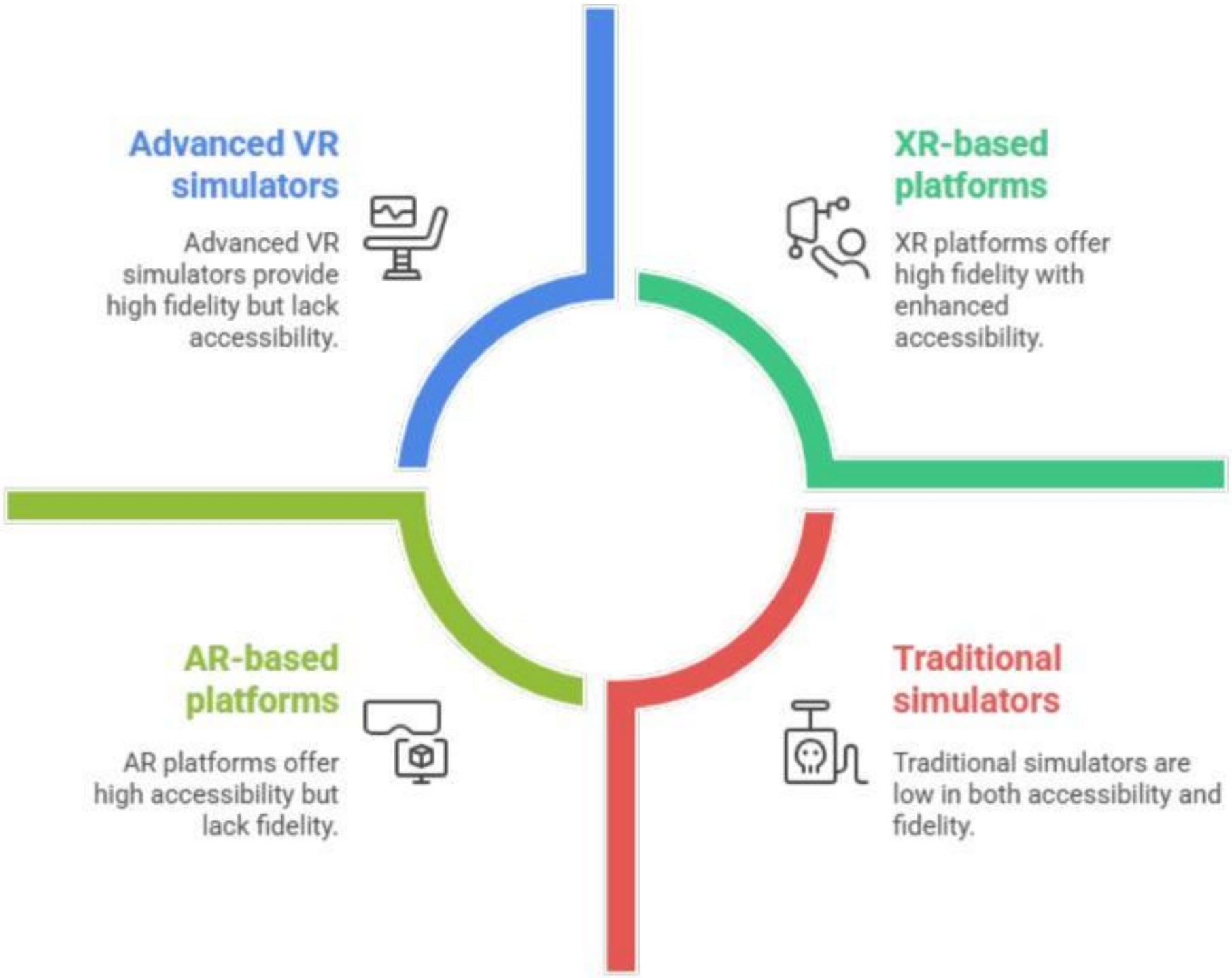

**Fig. 1** Comparative overview of simulation technologies used in surgical education. The figure contrasts traditional simulators, AR-based platforms, advanced VR simulators, and XR-based systems based on their fidelity and accessibility. XR-based platforms achieve a balance between high-fidelity and broad accessibility, representing the next evolution in immersive surgical training.

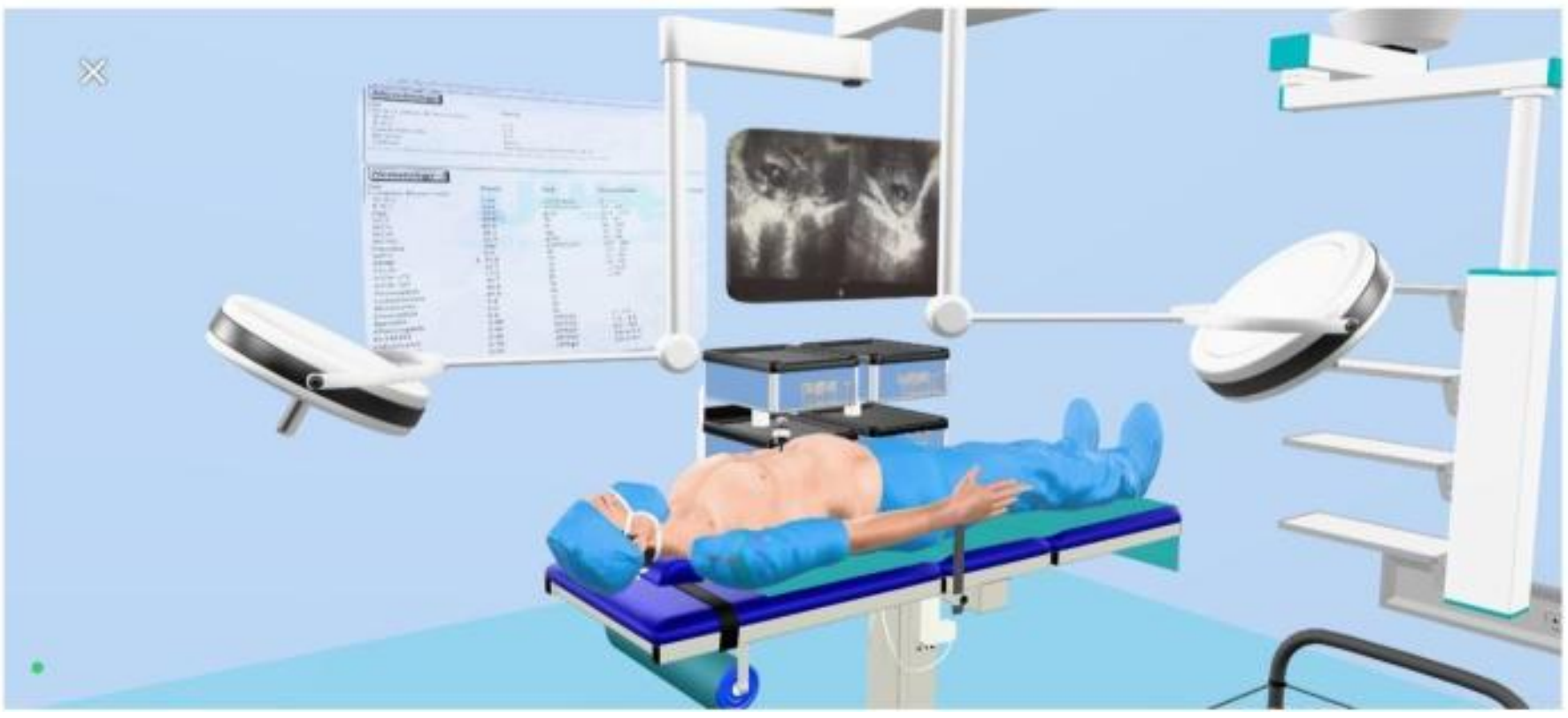

**Fig. 2** Environment of the XR platform

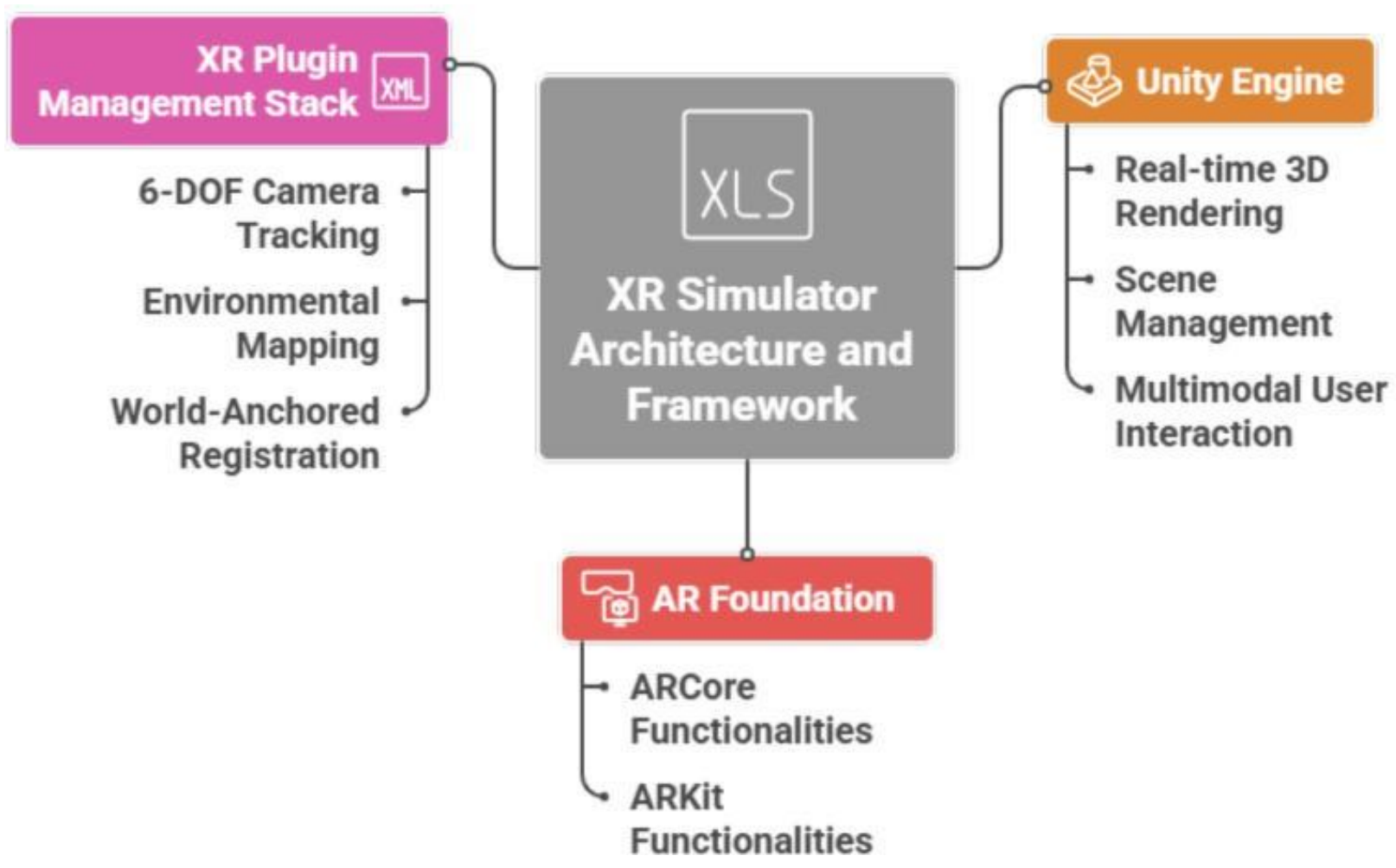

**Fig. 3** System architecture of the proposed XR-based laparoscopic training platform. The framework integrates the XR Plugin Management Stack for environmental tracking, AR Foundation for device compatibility through ARCore and ARKit, and the Unity Engine for real-time rendering and multimodal user interaction, forming a cohesive architecture for interactive simulation and feedback.

### 3.2 Virtual Tool Manipulation and Hand Interaction

Unlike traditional marker-based tracking systems, this framework employs an active hand-based interaction module for tool manipulation. Trainees manipulate the virtual trocar using touch-based gestures or spatial hand-tracking input. The interaction relies on a raycasting technique to map the user's input to the virtual scene. When the user initiates a grab or drag, a ray denoted as $R_{view}$ is cast from the camera, $C_{cam}$, to locate the interaction point, $P_{contact}$, on the patient's abdomen mesh:

$$R_{view}(t) = C_{cam} + t \cdot \vec{D}_{ray} \quad (1)$$

where $\vec{D}_{ray}$ is the normalized direction vector. Through constrained movement mapping, the trocar maintains anatomically appropriate contact with the abdominal model, simulating the physical fulcrum interaction described in MIS training literature.

### 3.3 Angular Deviation Estimation

Consistent with laparoscopic safety guidelines, the training emphasizes that the trocar insertion be perpendicular to the abdominal wall. The platform continuously evaluates the deviation angle θ between the trocar's insertion direction, $\vec{V}_{tool}$, and the ideal path, defined by the inverted surface normal, $-\vec{N}_{surface}$, at $P_{contact}$.

The deviation angle θ is computed using the dot product of the unit vectors:

$$\theta = \arccos(-\vec{V}_{tool} \cdot \vec{N}_{surface}) \quad (2$$

Where $\theta = 0^{\circ}$ signifies a perfect alignment. This model generates a quantitative,
objective measure that is scarce in many benchtop and low-cost

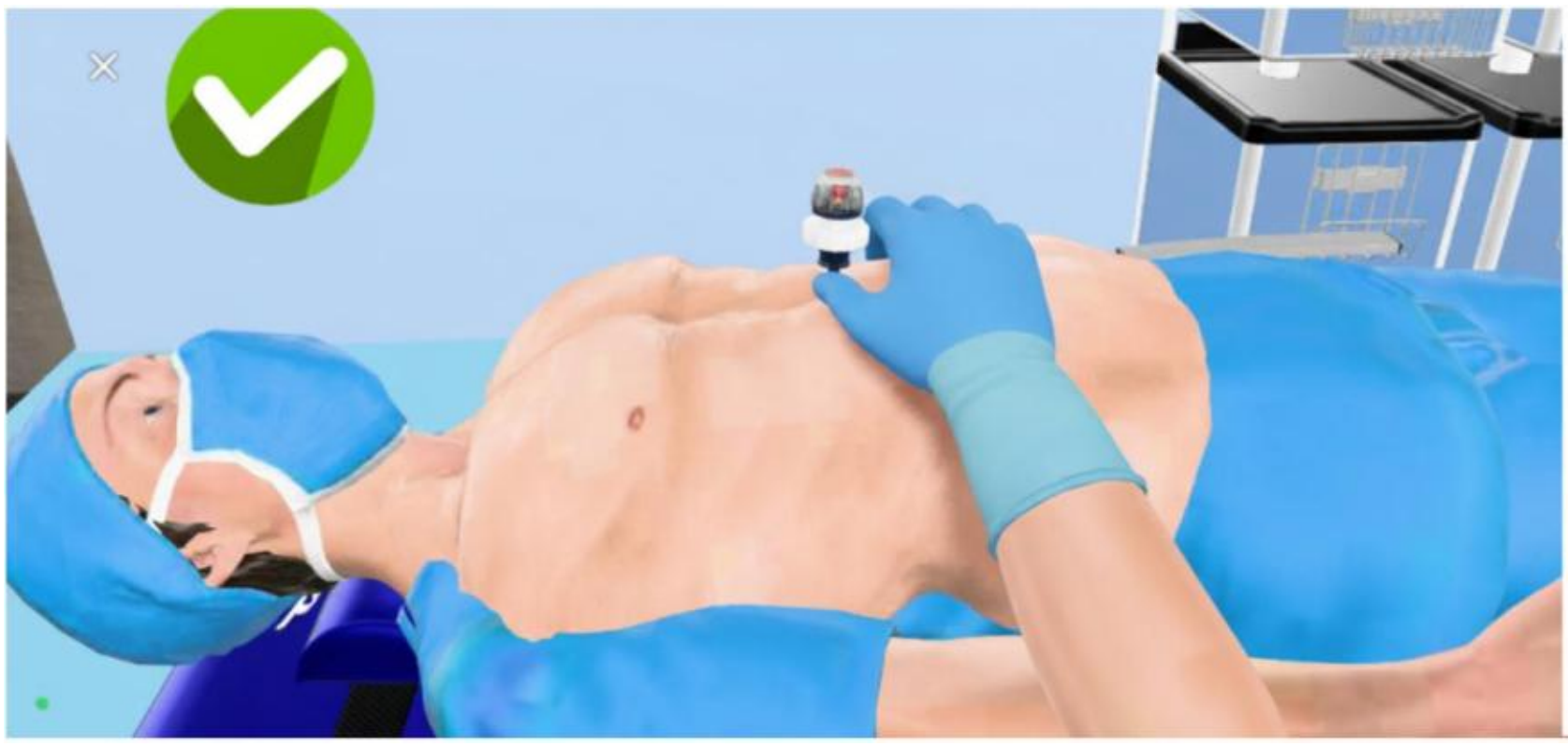

**Fig. 4** Visual feedback: when the angular deviation is within the threshold, $\epsilon$, which is set at $5^{\circ}$.

## 3.4 Gamified Assessment and Scoring

A gamified scoring system is implemented to provide immediate performance feedback. To convert the precision task into a structured learning loop, a gamified scoring algorithm rewards correct alignment and penalizes errors. A continuous function for score change, denoted as $\Delta S$, is defined based on a predefined tolerance threshold,$\epsilon$, which is set at $5^{\circ}$:

$$\Delta S = \begin{cases} +\alpha & \text{if } 0 \leq \theta \leq \epsilon \\ -\beta & \text{if } \quad \theta > \epsilon \end{cases} \tag{3}$$

Where $\alpha$ is the reward factor, and $\beta$ is the penalty factor. The total score, $S_{final}$, aggregates the instantaneous $\Delta S$ values over the session duration T.

## 3.5 Integrated Pseudo-Haptics

The platform integrates pseudo-haptic substitution to compensate for the absence of force-feedback hardware, emphasizing alerts through vibration patterns and visual contrast.

1. **Haptic Error Alert:** The device's vibration provides tactile warning. The haptic function, $H(\theta)$, triggers the alert only when angular error exceeds the tolerance threshold:

$$H(\theta) = \begin{cases} \text{Handheld.Vibrate ()} & \text{if } \theta > \epsilon \quad \text{(Error State)} \\ \text{No Action} & \text{if } \theta \leq \epsilon \quad \text{(Correct State)} \end{cases} \tag{4}$$

2. **Visual Feedback:** The material color of the virtual trocar $C_{mat}$ changes dynamically: Green for correct insertion ($\theta \leq \epsilon$) and Red for deviation ($\theta > \epsilon$), as shown in Fig 5

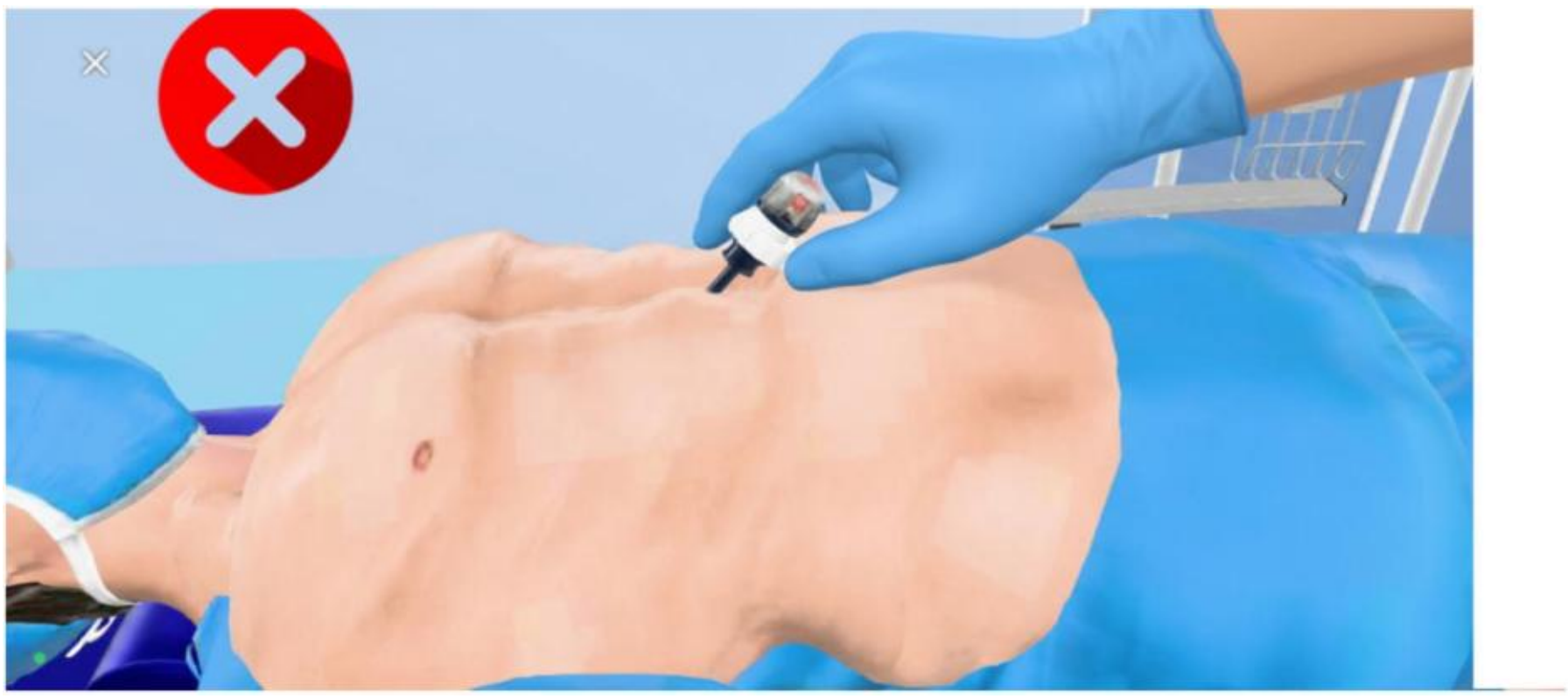

**Fig. 5** Visual feedback: when the angular deviation exceeds the threshold, $\epsilon$, which is set at $5^{\circ}$.

### 3.6 Medical Data Visualization Pipeline

A native gallery plugin enables trainees to import patient images such as ultrasound and lab reports. These assets are converted into Texture2D objects and dynamically applied to Quad-based 3D planes, which serve as virtual monitors spatially registered within the XR environment(Fig. 6). This situates the procedural training within a clinically contextualized environment, aligning with modern multimodal surgical skill acquisition approaches.

## 4 Experimental Setup

The empirical validation of the XR training platform involved a controlled study designed to assess the learning curve and efficacy of the gamified pseudo-haptic feedback mechanism. This section details the cohort characteristics, hardware/software specifications, and the standardized protocol used for data acquisition.

### 4.1 Participant Cohort

A total of $N = 20$ participants (medical students, mean age $22.0 \pm 2.5$ years) were recruited for the study. The inclusion criteria required that all participants had:

1. No prior clinical experience in laparoscopic surgery.
2. No previous exposure to high-fidelity surgical simulation trainers.
3. Normal or corrected-to-normal vision (essential for XR spatial tasks).

All participants provided informed consent prior to the commencement of the study. The study was approved by the institutional ethics review board.

### 4.2 Hardware and Software Configuration

The simulation was executed on a commercial, high-performance mobile device supporting the required XR capabilities (ARCore/ARKit). This device served as the primary interface for both visualization and interaction.

- **Development Platform:** Unity Engine (Version 2021.3 LTS).
- **XR Framework:** Unity AR Foundation package integrating ARCore and ARKit.
- **Input:** Custom C# scripts managing the hand-based interaction module and raycasting logic for virtual tool manipulation.
- **Feedback System:** Device's native vibration motor, controlled by the Handheld.Vibrate() function was utilized for the pseudo-haptic error alerts.
- **Data Logging:** All performance metrics, including the angular deviation ($\theta$), instantaneous score changes ($\Delta S$), and correction latency ($L_{corr}$), were automatically logged into a secure local database at a frequency of 60 Hz.

## 4.3 Experimental Protocol and Task Structure

The experimental procedure was standardized across all participants:

1. **Environment:** All sessions were conducted in a quiet, dedicated laboratory setting to minimize external distractions.
2. **Session Structure:** Each participant completed ten distinct training sessions over a period of two weeks, with a minimum of one day's rest between sessions to minimize fatigue effects.
3. **Task:** The core task for each session was the precise insertion of the virtual trocar into a designated target area on the virtual patient model, strictly adhering to the perpendicular constraint.
4. **Initial Calibration:** Before the first session, participants completed a 5-minute familiarization trial to adapt to the XR visual cues and the hand-based interaction mechanism.
5. **Data Collection:** The system autonomously initiated the gamified scoring and pseudo-haptic feedback loop upon task start and ceased recording when the insertion objective was met, or the session timer expired.

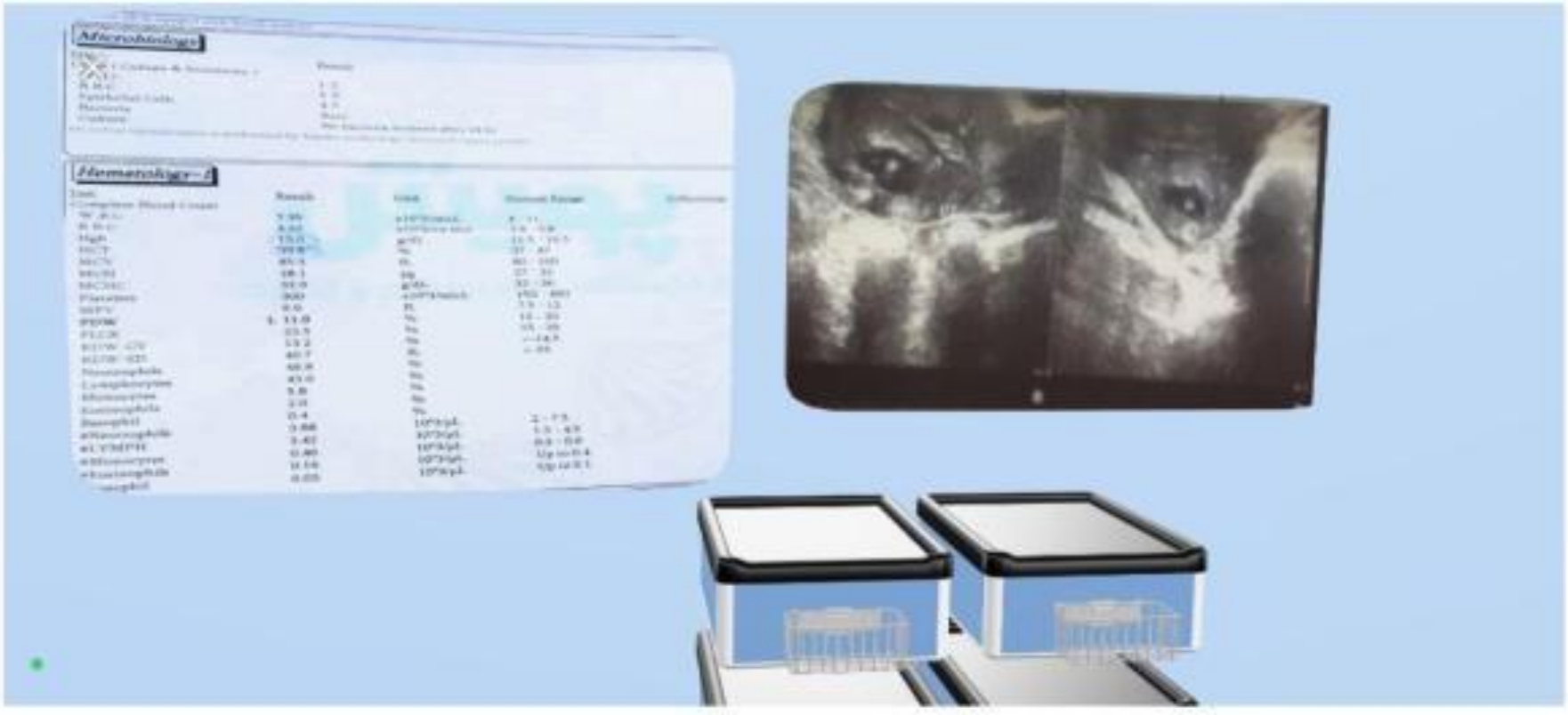

**Fig. 6** Patient data converted into Texture2D and applied to 3D planes.

## 5 Results

This section presents the empirical validation of the XR-based laparoscopic training system, focusing on its ability to accurately track virtual instrument manipulation, provide effective pseudo-haptic feedback, and enhance learning through gamified assessment. A controlled study was conducted with 20 medical students, mean age 22 ± 2.5 years, all novices in laparoscopic procedures. Each participant completed 10 sessions, with continuous logging of angle deviation, scoring, and haptic-response behaviors.

### 5.1 Angular Precision and Error Rate

The key objective of the training simulator is to minimize angular deviation ($\theta$) during trocar insertion. The framework sampled $\theta$ values at a rate of 60Hz during each insertion attempt. Angular Error Rate (AER) defines the percentage of time a trainee operates outside the permissible angular tolerance $\epsilon$ of $5^\circ$:

$$\text{AER} = \frac{\sum_{t=0}^{T_{\text{session}}} I(\theta(t) > \epsilon)}{T_{\text{session}} \cdot f} \times 100\% \tag{5}$$

Where $I(\cdot)$ is an indicator function–1 if true, 0 if false–$T_{\text{session}}$ is the duration of the session in seconds, and $f$ is the sampling frequency.

As displayed in Fig .7, the average angular deviation across all participants significantly decreased from an initial $\sim 15.2^\circ \pm 3.1^\circ$ (mean ± standard deviation) in Session 1 to $\sim 4.8^\circ \pm 1.2^\circ$ by Session 10 ($p < 0.001$). This strong learning trend confirms the framework's effectiveness. The AER similarly showed a marked reduction, dropping from an average of 68.5% in early sessions to 11.3% in the final sessions, as shown in Fig. 8.

These findings confirm the platform's capacity to support high-precision motor training through continuous monitoring and feedback.

### 5.2 Gamification Performance and Engagement

The integrated gamification module provides a quantitative measure of skill retention and reinforcement. The total score, $S_{\text{final}}$, at the end of each session, derived from the $\Delta S$ function, $\alpha = +10$ points for correct, $\beta = -5$ points for incorrect per second, serves as a direct indicator of overall performance.

Session scores were normalized into a Gamification Performance Index (GPI), scaled from 0 to 100, allowing for fair comparison across varying session lengths:

$$\text{GPI} = \frac{S_{\text{final}} - S_{\min}}{S_{\max} - S_{\min}} \times 100 \tag{6}$$

Where $S_{\min}$ and $S_{\max}$ are the theoretical minimum and maximum achievable scores for a given session duration, respectively.

The average GPI demonstrated a significant positive correlation with the number of training sessions, rising from an initial mean of 12.4% to a final mean of 88.9%, as seen in Fig.6. This rapid progression demonstrates that trainees responded strongly.

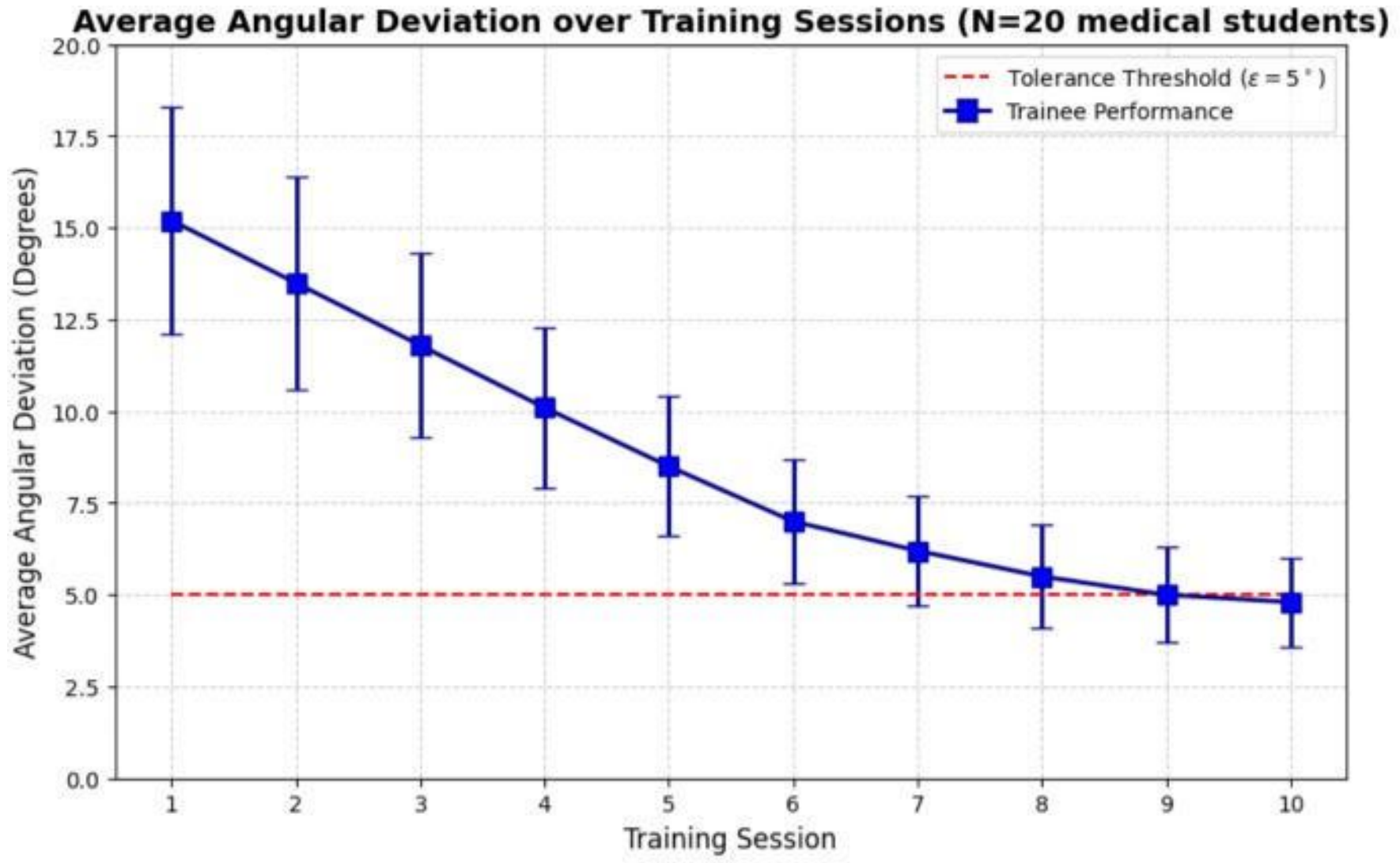

**Fig.7** Average angular deviation showed substantial decrease over 10 training sessions.

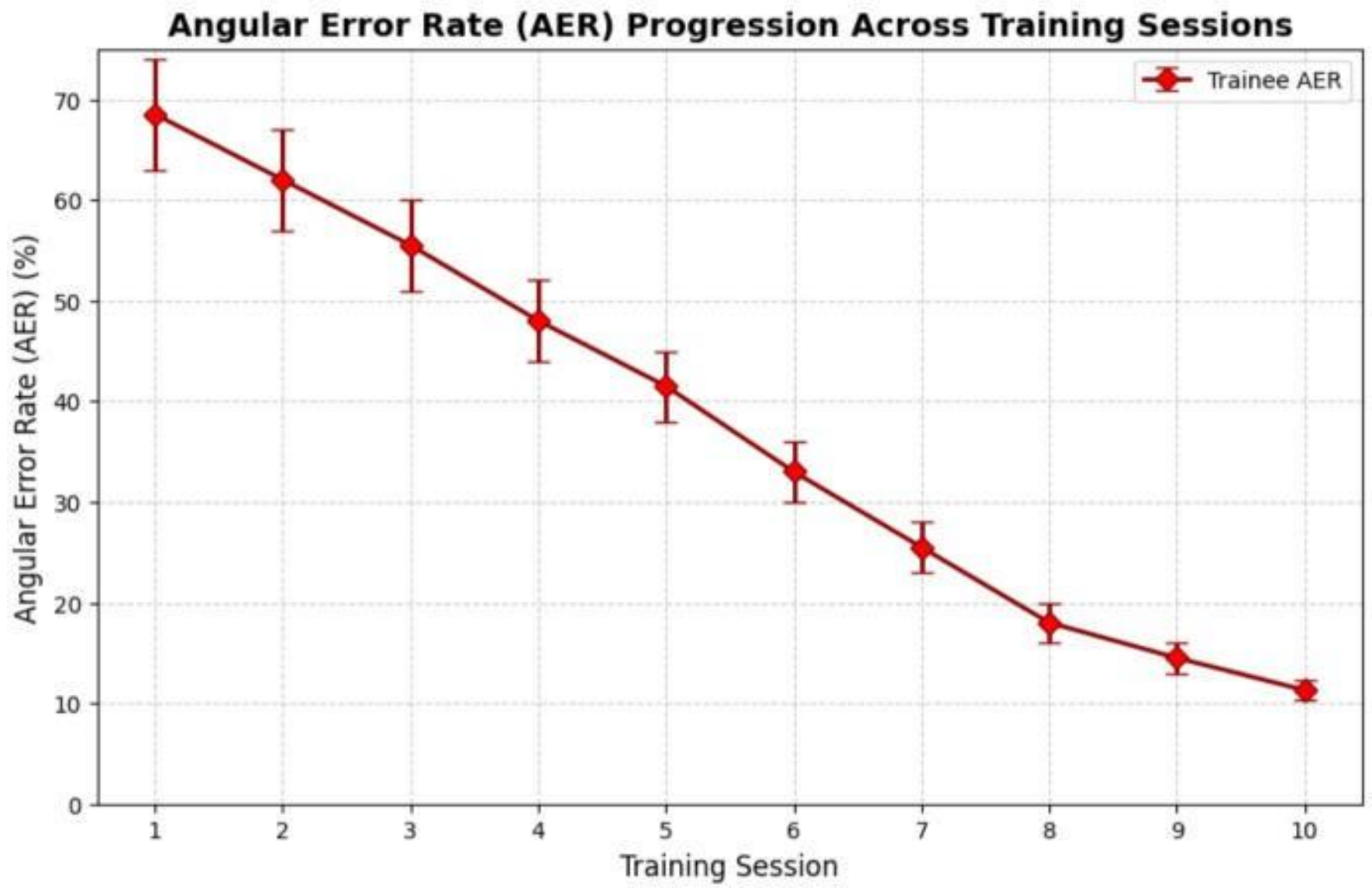

**Fig. 8.** Angular Error Rate (AER) showing a marked reduction across training sessions.

to the reward–penalty structure, with performance correlating directly to adherence to proper trocar alignment.

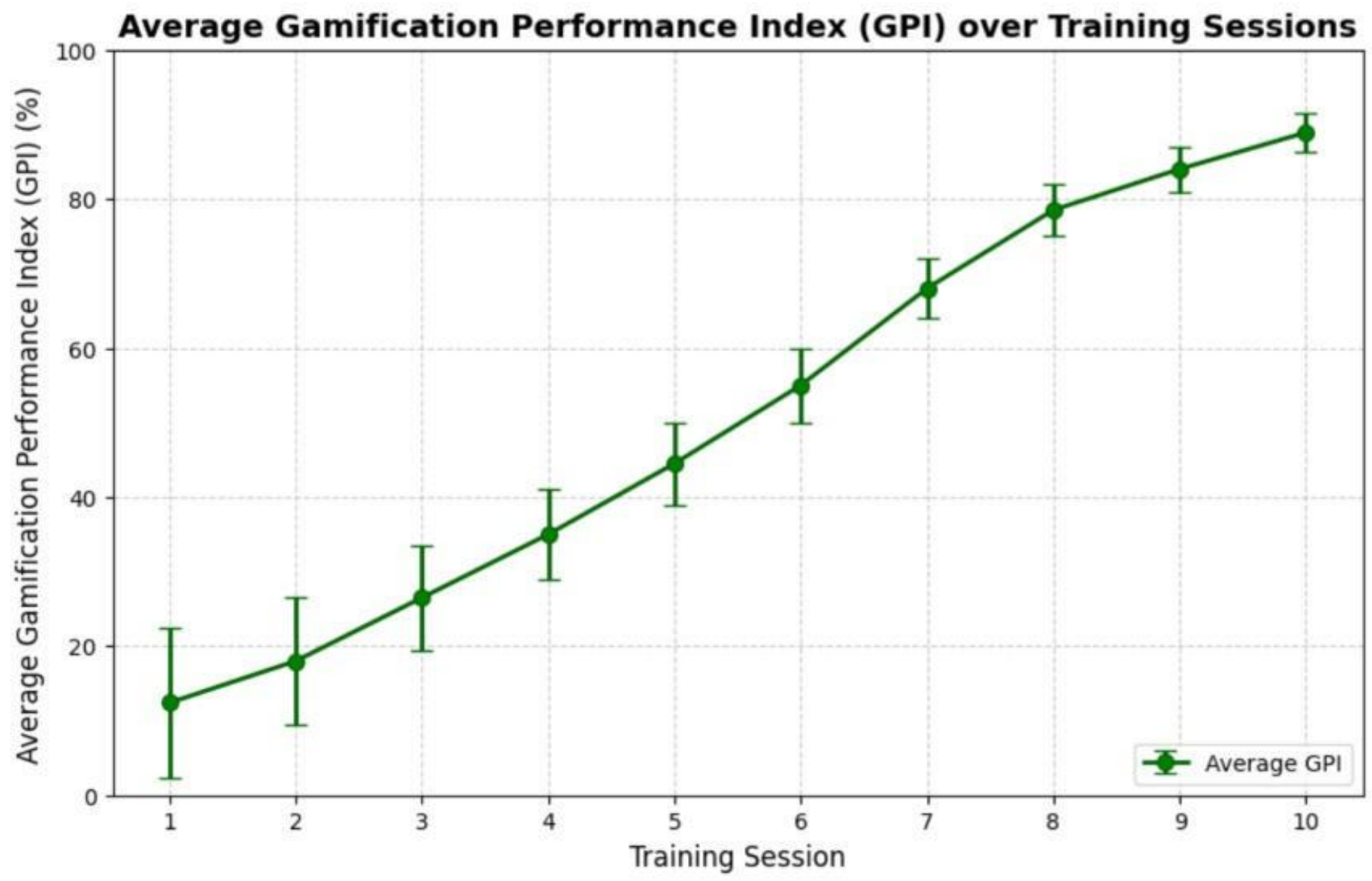

**Fig. 9** The average GPI showed a significant positive relationship with the number of training sessions.

## 5.3 Pseudo-Haptic Feedback Efficacy

The effectiveness of the pseudo-haptics mechanism in reinforcing proper technique was evaluated by analyzing the immediate trainee response to the vibration alert, where H(θ) = Vibrate().

The correction latency, the time elapsed between the trigger of the vibration alert and the moment the trainee brings the angle back into the acceptable range, was defined as:

$$L_{corr} = t_{corrected} - t_{alert} \quad (7)$$

The mean correction latency improved from $0.85 \pm 0.15$ seconds in Session 2 to $0.32 \pm 0.08$ seconds by Session 8. This reduction signifies that the dual haptic feedback mechanism effectively trained users to rapidly recognize and correct misalignment, allowing for faster development of proprioceptive awareness and demonstrating the successful implementation of the pseudo-haptic feedback loop.

# 6 DISCUSSION

The findings validate the core premise of this platform, demonstrating that an XR-based simulation environment integrated with a gamified assessment and pseudo-haptics can significantly accelerate the acquisition of precise technical skills. The proposed method directly addresses the primary challenges identified and specifically enhances the perpendicular trocar insertion technique fundamental in laparoscopic surgery.

### 6.1 XR Interaction as a Catalyst for Early Skill Acquisition

The shift from a marker-based AR system to an XR framework with direct hand-tracked manipulation represents a critical advancement. Conventional systems often confine interaction to pre-tracked physical tools, limiting the scope for smooth virtual manipulation. By incorporating a direct interaction model that mirrors real trocar handling, the platform achieved a high degree of interaction fidelity. This design incorporates intuitive instrument-hand mapping to reduce the MIS learning curve. The significant reduction in angular deviation confirms that mapping the trainee's hand movements directly to the virtual instrument, coupled with the immediate visual cues of the XR environment, forms a highly effective modality for surgical training. The continuous calculation of the angular deviation provided real-time, objective feedback that traditional non-digital training methods lack.

### 6.2 Gamification as a Structured Reinforcement Mechanism

The integration of the real-time scoring and the Gamification Performance Index transformed practice into a goal-oriented, repeatable task, helping increase trainee motivation. The steep rise in GPI shows that the defined reward and penalty structure encouraged repeated practice and improved attention to improper tool handling. The immediate feedback loop mechanism addresses a major limitation in surgical training, where feedback is often subjective and intermittent.

### 6.3 Pseudo-Haptics as a Practical Alternative to Force Feedback

Although high-end VR simulators provide true force feedback, they lack wide accessibility due to financial and geographical constraints. The substantial decrease in Correction Latency indicates that the multi-sensory pseudo-haptics effectively bridge the sensory gap left by the absence of specialized force-feedback hardware. The dual-haptic feedback acted as a powerful, low-latency warning, enabling trainees to self-correct the angle before significant score penalties were incurred. This highlights the feasibility of accessible, cost-effective, scalable pseudo-haptic solutions to replace expensive physical simulators in early-stage skill acquisition.

### 6.4 Implications for Training Scalability

The combination of resource-efficient hardware, XR overlays, pseudo-haptics, and gamification forms a training ecosystem that is:

1. Accessible: No dedicated simulation center required,
2. Quantitative: Objective metrics replacing subjective feedback,
3. High-frequency: enabling repeated practice,
4. Engaging: gamified structure sustains motivation.

## 7 Conclusion

This work introduced and validated a novel, accessible XR platform designed to enhance fundamental laparoscopic skills through instrument tracking, gamified assessment, and pseudo-haptic cues. By focusing on the clinically essential task of trocar alignment and angular control, the framework offers an immersive, portable, and resource-efficient alternative to high-end laparoscopic simulators. These features enable continuous deliberate practice outside traditional training centers.

The platform's core contribution lies in the synergistic integration of cost-effective hardware with sophisticated feedback mechanisms. Continuous tracking of the instru-ment's deviation from the ideal perpendicular vector provides objective, real-time performance metrics. This quantitative assessment is tightly coupled with a mathematically defined gamification layer, transforming repetitive motor practice into a structured, goal-oriented challenge that enhances engagement and promotes sustained skill retention. The pseudo-haptic feedback module further strengthens the learning loop by delivering subtle vibration alerts proportional to angular deviation magnitude. This module effectively reintroduces a non-visual sensory channel that reduces cognitive load and facilitates immediate error correction.

The proposed method successfully achieved its primary objective: participants demonstrated a substantial improvement in angular precision, as a significant reduction in average deviation was observed across ten sessions. The dramatic rise in the Gamification Performance Index and the marked reduction in correction latency confirm the effectiveness of the integrated multi-sensory design. Together, these findings provide strong evidence that accessible XR-based training can meaningfully accelerate early psychomotor learning in laparoscopy.

Overall, this work advances the state of resource-efficient surgical simulation by demonstrating that consumer-grade XR, when combined with objective metrics, gamified reinforcement, and pseudo-haptic cues, can replicate critical aspects of high-fidelity training. Future directions include expanding the assessment framework to more complex tasks such as tissue manipulation, cutting, and suturing. By prioritizing portability, measurability, and learner engagement, this XR framework represents a practical step toward democratizing high-quality laparoscopic training on a global scale.

## Declarations

**Funding:** The authors received no specific funding for this work.

**Conflicts of Interest:** The authors declare that they have no competing interests.

**Ethical Approval:** The study was approved by the institutional ethics review board.

**Informed Consent:** All participants provided informed consent prior to the commencement of the study.

**Affiliations:** The authors declare no institutional affiliations.